\documentclass[sigconf,natbib=true]{acmart}
\AtBeginDocument{%
  }

\usepackage{booktabs}
\usepackage{multirow}
\usepackage{colortbl}
\usepackage{xcolor}
\usepackage{makecell}
\usepackage{float}
\usepackage{enumitem}
\usepackage{amsmath}
\usepackage{twemojis}
\usepackage{algorithm}     
\usepackage{algpseudocode} 
\usepackage{bbm}
\usepackage{bm}

\newtheorem{theorem}{Theorem}[section]

\newtheorem{proposition}[theorem]{Proposition}
\newtheorem{corollary}[theorem]{Corollary}

\theoremstyle{definition}

\newtheorem{assumption}[theorem]{Assumption}

\theoremstyle{remark}

\newcommand{\imps}[1]{\textcolor{red}{\textbf{+#1\%*}}}  
\newcommand{\imp}[1]{\textcolor{red}{\textbf{+#1\%}}}     
\newcommand{\impn}[1]{-#1\%}                               

\setcopyright{acmlicensed}
\copyrightyear{2018}
\acmYear{2018}
\acmDOI{XXXXXXX.XXXXXXX}
\acmConference[Conference acronym 'XX]{Make sure to enter the correct
  conference title from your rights confirmation email}{June 03--05,
  2018}{Woodstock, NY}
\acmISBN{978-1-4503-XXXX-X/2018/06}

\begin{document}

\title{Parallel Latent Reasoning for Sequential Recommendation}

\author{Jiakai Tang}
\authornote{Work done during an internship at Alibaba Group.}
\email{tangjiakai5704@ruc.edu.cn}
\affiliation{%
  \institution{GSAI, Renmin University of China}
  \city{Beijing}
  \country{China}
}

\author{Xu Chen}
\authornote{Both authors are corresponding authors.}
\email{xu.chen@ruc.edu.cn}
\affiliation{%
    \institution{GSAI, Renmin University of China}
    \city{Beijing}
    \country{China}
}

\author{Wen Chen}
\authornotemark[2]
\email{chenyu.cw@alibaba-inc.com}
\affiliation{%
    \institution{Alibaba Group}
    \city{Beijing}
    \country{China}
}

\author{Jian Wu}
\email{joshuawu.wujian@alibaba-inc.com}
\affiliation{%
    \institution{Alibaba Group}
    \city{Beijing}
    \country{China}
}

\author{Yuning Jiang}
\email{mengzhu.jyn@alibaba-inc.com}
\affiliation{%
    \institution{Alibaba Group}
    \city{Beijing}
    \country{China}
}

\author{Bo Zheng}
\email{bozheng@alibaba-inc.com}
\affiliation{%
    \institution{Alibaba Group}
    \city{Beijing}
    \country{China}
}

\renewcommand{\shortauthors}{Trovato et al.}

\begin{abstract}

Capturing complex user preferences from sparse behavioral sequences remains a fundamental challenge in sequential recommendation. Recent latent reasoning methods have shown promise by extending test-time computation through multi-step reasoning, yet they exclusively rely on depth-level scaling along a single trajectory, suffering from diminishing returns as reasoning depth increases. To address this limitation, we propose \textbf{Parallel Latent Reasoning (PLR)}, a novel framework that pioneers width-level computational scaling by exploring multiple diverse reasoning trajectories simultaneously. PLR constructs parallel reasoning streams through learnable trigger tokens in continuous latent space, preserves diversity across streams via global reasoning regularization, and adaptively synthesizes multi-stream outputs through mixture-of-reasoning-streams aggregation. Extensive experiments on three real-world datasets demonstrate that PLR substantially outperforms state-of-the-art baselines while maintaining real-time inference efficiency. Theoretical analysis further validates the effectiveness of parallel reasoning in improving generalization capability. Our work opens new avenues for enhancing reasoning capacity in sequential recommendation beyond existing depth scaling. 

\end{abstract}




\maketitle

\section{Introduction}

Recommender systems have become indispensable infrastructure in modern digital ecosystems, facilitating personalized content discovery and user engagement across diverse domains such as e-commerce~\cite{tang2025interactive,wang2025reaseq}, streaming platforms~\cite{guo2019streaming,wang2018streaming}, and social networks~\cite{konstas2009social,yang2012circle}. However, a persistent challenge in recommendation lies in the sparsity of user behavioral data, which fundamentally limits traditional models' capability to perform complex logical inference for uncovering users' potential interest patterns and preference dynamics~\cite{tang2024robust,tang2024towards,yu2020category}. Inspired by recent breakthroughs in large language models (LLMs)~\cite{huang2023towards,plaat2024reasoning,chen2025towards}, where incorporating reasoning mechanisms has demonstrated remarkable improvements in challenging tasks such as mathematical problem-solving~\cite{wang2025survey,ahn2024large} and code generation~\cite{li2023chain,ding2024semcoder}, an emerging paradigm in sequential recommendation has begun to embrace \emph{test-time computation} strategies~\cite{zhang2025towards}. This paradigm shift introduces reasoning capabilities---both explicit and latent---that transform the conventional shallow pattern-matching approach (input-output) into a sophisticated reasoning-driven framework (input-think-output), substantially enhancing models's computational power. 

Despite their promise, existing reasoning-enhanced recommendation methods encounter several critical limitations. For explicit reasoning approaches, two fundamental challenges persist: \textbf{(1) Inference latency overhead}: The generation of verbose reasoning chains incurs substantial computational costs, creating prohibitive delays for real-time online serving scenarios. \textbf{(2) Reasoning chain definition ambiguity}: Unlike mathematical reasoning or code synthesis tasks, where LLMs (such as DeepSeek-R1~\cite{guo2025deepseek}, Qwen3~\cite{yang2025qwen3}, and Kimi1.5~\cite{team2025kimi}) leverage extensive Chain-of-Thought (CoT) data during post-training that can be \textit{rigorously verified against objective ground truth}. However, recommendation scenarios lack \textit{well-defined} reasoning trajectories. The absence of expert-annotated, high-quality ``correct'' reasoning chains makes it inherently challenging to guide explicit reasoning in recommendation contexts.

In contrast, latent reasoning-enhanced recommendation methods adopt a data-driven, annotation-free approach to unlock models' reasoning potential. However, these methods face a distinct bottleneck: current techniques mainly rely on \emph{depth computational scaling}---deepening the reasoning process through recurrent autogressive steps. Empirical evidence suggests that further increases in reasoning depth yield diminishing or even negative performance gains~\cite{tang2025think,zhang2025reinforced,liu2025lares}. We argue this limitation likely stems from two intertwined factors: \textbf{(i)} the model's initial reasoning direction may be suboptimal, and \textbf{(ii)} error accumulation along extended reasoning chains progressively degrades the thinking quality, ultimately constraining the model's reasoning capability ceiling.

In this work, to meet the demands of real-time online inference, we adopt the latent reasoning paradigm while rethinking its computational approach. Rather than pursuing deeper reasoning along a singular trajectory, we draw inspiration from cognitive science~\cite{clark1989microcognition,jackendoff2011parallel}, which reveals that humans typically explore \emph{multiple plausible solution paths concurrently} before converging to a comprehensive conclusion. Building upon this insight, we propose a novel computational scaling dimension: \textbf{\emph{width-level reasoning enhancement}}. Multi-stream latent reasoning enables the model to capture users' dynamic and multifaceted interest preferences through diverse reasoning trajectories, preventing premature lock-in to a single, potentially suboptimal thought process. 
However, applying this idea is non-trivial due to the following challenges:
\textbf{Challenge 1: How to construct multi-stream reasoning?} Unlike discrete decoding in LLMs, where techniques such as beam search~\cite{xie2023self,wiseman2016sequence}, self-consistency~\cite{wang2025a2r,aggarwal2023let}, and majority voting~\cite{wang2022self,chen2024more} operate on language token sequences, it remains unclear how to effectively construct multiple reasoning streams in the continuous latent representation space. 
\textbf{Challenge 2: How to avoid reasoning homogeneity?} While width scaling increases computational capacity, it is critical to prevent multiple streams from converging to similar reasoning patterns, leading to redundant resource consumption without bringing positive benefits.
\textbf{Challenge 3: How to aggregate different reasoning streams?} After parallel exploration, the model needs to synthesize insights from multiple reasoning pathways into a unified solution. Naive aggregation strategies (e.g., mean pooling) risk diluting the contributions of superior reasoning streams while amplifying the interference from inferior ones. 

To address these challenges, we propose a simple yet effective framework for parallel latent reasoning in sequential recommendation, named \textbf{Parallel Latent Reasoning (PLR)}. Specifically, to expand the latent representation from a sequential encoder to multiple streams, we introduce learnable \emph{trigger tokens} that actively guide the model to conduct parallel multi-stream reasoning. This simple design effectively resolves the challenge of stream construction in continuous latent space. Furthermore, to enhance the distinctiveness among different streams and across reasoning steps within each stream, we introduce a \emph{global reasoning regularization} mechanism that mitigates the homogeneity issue. For aggregating results from multiple parallel reasoning streams, we design an \emph{mixture-of-reasoning-streams} aggregation module that adaptively weights and combines outputs from different streams to synthesize the final sequence representation. Additionally, to improve the model's reasoning robustness, we propose a \emph{reasoning contrastive learning} objective that enhances the model's ability to handle sparse user behaviors. We also provide in-depth theoretical analysis to explain the effectiveness of our approach, offering insights that advance the understanding of reasoning-enhanced recommendation.

Our main contributions are summarized as follows:
\begin{itemize}[leftmargin=*,noitemsep,topsep=0pt]
    \item We pioneer the exploration of \emph{width-level computational scaling} for latent reasoning in sequential recommendation, enabling a novel architecture that synergizes breadth and depth reasoning while maintaining real-time inference efficiency.
    \item We propose PLR, a model-agnostic framework for parallel latent reasoning that employs learnable trigger tokens to construct diverse reasoning streams in continuous latent space, regularizes global reasoning patterns to preserve diversity, and aggregates multi-stream outputs through mixture-of-reasoning-streams.
   \item Extensive experiments on three real-world datasets demonstrate that PLR achieves substantial improvements and establishes new ceilings for sequential recommendation, opening new avenues for latent reasoning-enhanced recommendation research.
\end{itemize}

\section{Preliminary}
\subsection{Problem Formulation}

Let $\mathcal{U} = \{u_1, u_2, \ldots, u_{|\mathcal{U}|}\}$ denote the user set and $\mathcal{V} = \{v_1, v_2, \ldots, v_{|\mathcal{V}|}\}$ denote the set of items. For each user $u \in \mathcal{U}$, we denote their chronologically ordered interaction sequence as $\mathcal{S}_u = [v_1^u, v_2^u, \ldots, v_n^u]$, where $v_i^u \in \mathcal{V}$ represents the $i$-th item interacted by user $u$, and $n$ is the sequence length. The task of sequential recommendation is to predict the next item $v_{n+1}^u$ that user $u$ is most likely to interact with, given their historical sequence $\mathcal{S}_u$. 
Formally, this problem can be formulated as learning a function $f_\theta: \mathcal{S}_u \rightarrow \mathcal{V}$ that maximizes the conditional probability of the ground-truth target item:
\begin{equation}\label{eq:next-item-prediction}
    \theta^* = \arg\max_{\theta} \sum_{u \in \mathcal{U}} \log p(v_{n+1}^u | \mathcal{S}_u; \theta),
\end{equation}
where $\theta$ denotes the model parameters. During inference, the model ranks all candidate items based on their predicted probabilities and recommends the top-$K$ items with the highest scores.

\subsection{Latent Reasoning-Enhanced Recommendation}

Traditional recommendation methods rely on a single forward pass to encode the sequence representation, where the fixed computational budget limits their modeling capability, particularly in sparse interaction scenarios. To enable more fine-grained reasoning and deeper representation learning, recent works have explored latent reasoning-enhanced recommender systems~\cite{tang2025think,liu2025lares,dai2025onepiece,zhang2025slow}, which shift from the conventional shallow \emph{input-output} paradigm to a multi-step \emph{input-think-output} reasoning framework.

Formally, given an input sequence $\mathcal{S}_u$, a latent reasoning-enhanced model first encodes the sequence into a latent representation $\mathbf{h}_0$ through an encoder $f_{\text{enc}}$:
\begin{equation*}
    \mathbf{h}_0 = f_{\text{enc}}(\mathcal{S}_u; \theta_{\text{enc}}),
\end{equation*}
where $\theta_{\text{enc}}$ denotes the encoder parameters. Subsequently, the model performs $T$ iterative reasoning steps to progressively refine the representation:
\begin{equation*}
    \mathbf{h}_t = f_{\text{rea}}(\mathbf{h}_{t-1}; \theta_{\text{rea}}), \quad t = 1, 2, \ldots, T,
\end{equation*}
where $f_{\text{rea}}$ is the reasoning module with parameters $\theta_{\text{rea}}$ (which can be shared with or distinct from $\theta_{\text{enc}}$ depending on the specific method). Finally, the model generates the prediction based on the refined representation $\mathbf{h}_T$. The core principle of this paradigm is to enhance the model's computational expressiveness through iterative reasoning, thereby improving accuracy on complex tasks.

However, empirical studies on existing latent reasoning-enhanced methods reveal a critical limitation: after stacking multiple reasoning steps (e.g., 2 steps in ReaRec~\cite{tang2025think}, 3-4 steps in LARES~\cite{liu2025lares}, and 1 step in LatentR$^3$~\cite{zhang2025reinforced}), models often exhibit marginal or even negative performance gains, a phenomenon known as \emph{over-thinking}~\cite{peng2025revisiting,yadav2025hop}. This issue suggests that depth-only computational scaling may be insufficient to unlock the model's full potential. More advanced reasoning mechanisms are needed to further unleash the latent reasoning capabilities of recommendation models.

\section{Methodology}

\begin{figure}
    \centering
    \includegraphics[width=\linewidth]{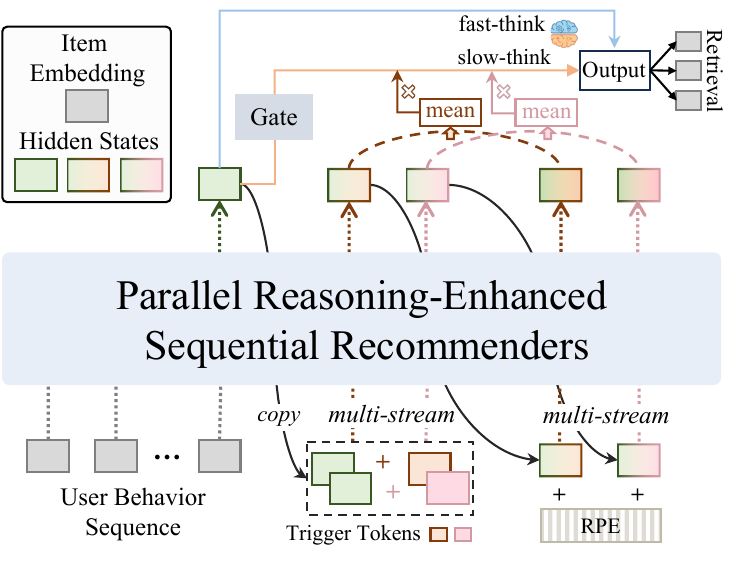}
    \vspace{-20pt}
    \caption{Overall architecture of the Parallel Latent Reasoning framework. RPE denotes reasoning position embedding.}
    \label{fig:framework}
    \vspace{-10pt}
\end{figure}

In this section, we present the \textbf{Parallel Latent Reasoning (PLR)} framework for sequential recommendation. The overall architecture is illustrated in Figure~\ref{fig:framework}. We first introduce the parallel latent reasoning backbone in Section~\ref{sec:plr-backbone}, which extends current depth-only reasoning to width-level parallel streams through learnable trigger tokens. Subsequently, Section~\ref{sec:optimization} presents three key optimization mechanisms: global reasoning regularization to preserve diversity across reasoning streams, reasoning contrastive learning to enhance reasoning robustness, and mixture-of-reasoning-streams aggregation to adaptively synthesize multi-stream outputs. Finally, Section~\ref{sec:training-objective} discusses the complete training objective and the dual-process inference mechanism.

\subsection{Parallel Latent Reasoning Backbone}\label{sec:plr-backbone}

In this section, we introduce the core architecture of our Parallel Latent Reasoning (PLR) framework. We begin by describing the foundational components: the attention-based sequence encoder, depth-level latent reasoning, and our proposed width-level parallel reasoning approach.

\subsubsection{\textbf{Attention-based Sequence Encoding}}

Following mainstream sequential recommendation methods~\cite{wu2020sse,zivic2024scaling,zhang2024scaling}, we adopt the Transformer architecture~\cite{vaswani2017attention} as our backbone. Given a user interaction sequence $\mathcal{S}_u = [v_1^u, v_2^u, \ldots, v_n^u]$, we first embed each item into a dense vector representation. Formally, let $\mathbf{E} \in \mathbb{R}^{|\mathcal{V}| \times d}$ denote the item embedding matrix, where $d$ is the embedding dimension. The initial sequence representation is obtained by:
\begin{equation*}
    \mathbf{X}_0 = [\mathbf{e}_1, \mathbf{e}_2, \ldots, \mathbf{e}_n] + \mathbf{P},
\end{equation*}
where $\mathbf{e}_i = \mathbf{E}[v_i^u] \in \mathbb{R}^d$ is the embedding of item $v_i^u$, and $\mathbf{P} \in \mathbb{R}^{n \times d}$ represents the positional encoding that injects sequential order information into the model.
Subsequently, $\mathbf{X}_0$ is fed into a multi-layer multi-head self-attention module to capture complex dependencies among items in the sequence. Specifically, for the $l$-th layer, the computation is formulated as:
\begin{equation*}
    \mathbf{X}_l = \text{MultiHead}(\mathbf{X}_{l-1}) + \mathbf{X}_{l-1},
\end{equation*}
where the multi-head self-attention operation is defined as:
\begin{equation*}
    \text{MultiHead}(\mathbf{X}) = \text{Concat}(\text{head}_1, \ldots, \text{head}_H)\mathbf{W}^O,
\end{equation*}
with each attention head computed as:
\begin{equation*}
    \text{head}_h = \text{Attention}(\mathbf{X}\mathbf{W}_h^Q, \mathbf{X}\mathbf{W}_h^K, \mathbf{X}\mathbf{W}_h^V),
\end{equation*}
where $\mathbf{W}_h^Q, \mathbf{W}_h^K, \mathbf{W}_h^V \in \mathbb{R}^{d \times d_h}$ are learnable projection matrices for the $h$-th head, $\mathbf{W}^O \in \mathbb{R}^{Hd_h \times d}$ is the output projection matrix, and $H$ is the number of attention heads. After $L$ encoder layers, we obtain the final encoded representation $\mathbf{H}_{\text{enc}} = \mathbf{X}_L \in \mathbb{R}^{n \times d}$.

\subsubsection{\textbf{Depth-Level Latent Reasoning}}

Following recent latent reasoning methods (e.g., ReaRec~\cite{tang2025think}), we extend the conventional one-pass encoding paradigm to incorporate depth-level multi-step reasoning computation. Rather than directly using the encoder output $\mathbf{H}_{\text{enc}}$ as the final sequence representation, which is constrained by limited computational capacity for capturing fine-grained user interests, we introduce additional autoregressive reasoning steps to progressively refine the latent representation.

Specifically, we extract the representation of the last item in the sequence as the initial reasoning state: $\mathbf{h}_0 = \mathbf{H}_{\text{enc}}[-1] \in \mathbb{R}^d$. Then, the model performs $T$ iterative reasoning steps:
\begin{equation}
    \mathbf{h}_t = f(\mathbf{h}_{t-1}, \mathbf{r}_t; \theta), \quad t = 1, 2, \ldots, T,
\end{equation}
where $f$ denotes the reasoning module implemented as Transformer layers with shared parameters $\theta$ from the encoder, and $\mathbf{r}_t \in \mathbb{R}^d$ is the Reasoning Position Embedding (RPE) for step $t$ that helps distinguish the reasoning phase from the encoding phase.

\subsubsection{\textbf{Width-Level Parallel Reasoning}}

However, depth-only computational scaling remains susceptible to several critical limitations: suboptimal initial reasoning directions and error accumulation across continuous reasoning chains. To mitigate these challenges, we explore a novel dimension of computational scaling: \emph{width-level parallel latent reasoning}. The key challenge lies in extending the continuous latent state vector into multiple parallel streams in the latent space.

To enable this extension, we introduce \textbf{learnable trigger tokens} $\{\boldsymbol{\tau}_1, \boldsymbol{\tau}_2, \ldots, \boldsymbol{\tau}_M\} \in \mathbb{R}^d$ to explicitly guide different reasoning streams, where $M$ denotes the number of parallel reasoning streams. Specifically, we adopt a simplistic approach: we add each trigger token to the initial reasoning state $\mathbf{h}_0$ to obtain $M$ distinct beginning reasoning states, which is formulated as:
\begin{equation}
    \mathbf{h}_{0,m} = \mathbf{h}_0+ \boldsymbol{\tau}_m, \quad m = 1, 2, \ldots, M.
\end{equation}
This design allows each stream to start from a perturbed initial state, encouraging the exploration of diverse reasoning trajectories.

For the attention mechanism across multiple reasoning streams, we adopt a \textbf{stream-isolated causal reasoning} approach with shared input attention. Formally, for each stream $m$ at reasoning step $t$, the computation is formulated as:
\begin{equation}
    \mathbf{h}_{t,m} = \text{Attention}(\mathbf{Q}_{t,m}, \mathbf{K}_{t,m}, \mathbf{V}_{t,m}) + \mathbf{h}_{t-1,m},
\end{equation}
where:
\begin{equation}
\begin{aligned}
    \mathbf{Q}_{t,m} &= (\mathbf{h}_{t-1,m} + \mathbf{r}_t)\mathbf{W}^Q, \\
    \mathbf{K}_{t,m} &= [\mathbf{X}_0; \mathbf{h}_{1,m}; \ldots; \mathbf{h}_{t-1,m}]\mathbf{W}^K, \\
    \mathbf{V}_{t,m} &= [\mathbf{X}_0; \mathbf{h}_{1,m}; \ldots; \mathbf{h}_{t-1,m}]\mathbf{W}^V,
\end{aligned}
\end{equation}
here, $[\cdot; \cdot]$ denotes concatenation along the sequence dimension. This design ensures that: \textbf{(i)} each stream attends to the shared interaction context, preserving access to the original input information; \textbf{(ii)} reasoning within each stream follows a causal pattern, where the state at step $t$ only attends to previous steps within the same stream; \textbf{(iii)} different streams remain isolated during reasoning, enabling independent exploration of diverse reasoning paths.

After $T$ reasoning steps, we obtain $M$ refined representation groups $\{\mathbf{H}_{1}^{\text{rea}}, \mathbf{H}_{2}^{\text{rea}}, \ldots, \mathbf{H}_{M}^{\text{rea}}\}$, where $\mathbf{H}_{m}^{\text{rea}} = \{\mathbf{h}_{1,m}, \mathbf{h}_{2,m}, \ldots, \mathbf{h}_{T,m}\}$ corresponds to the reasoning trajectory of the $m$-th stream. Following the \textit{Ensemble Reasoning Learning (ERL)} strategy from ReaRec, we apply average pooling over the multi-step outputs within each reasoning stream to obtain the stream-specific representation:
\begin{equation}\label{eq:stream_output}
    \mathbf{z}_m = \frac{1}{T}\sum_{t=1}^{T}\mathbf{h}_{t,m}, \quad m = 1, 2, \ldots, M.
\end{equation}
The final sequence representation is then derived by aggregating the outputs $\{\mathbf{z}_1, \mathbf{z}_2, \ldots, \mathbf{z}_M\}$ from all reasoning streams. In Section~\ref{sec:aggregation}, we will introduce our adaptive mixture-of-reasoning-streams approach for combining multi-stream reasoning results.

\subsection{Multi-Stream Reasoning Optimization}\label{sec:optimization}

In this section, we introduce three key mechanisms to optimize the parallel reasoning process: global reasoning regularization to preserve diversity, reasoning contrastive learning to enhance robustness, and mixture-of-reasoning-streams for adaptive aggregation.

\subsubsection{\textbf{Global Reasoning Regularization}}

To prevent multiple reasoning streams from collapsing into homogeneous patterns, where different streams exhibit highly similar reasoning states, we introduce a \emph{global reasoning regularization} mechanism. Specifically, we enforce diversity constraints across all reasoning states (both within and across streams) through bidirectional Kullback-Leibler (KL) divergence regularization.

Formally, let $\mathbf{h}_{t,m}$ denote the hidden state at step $t$ of stream $m$. We compute the distribution over the item vocabulary for each reasoning state as:
\begin{equation*}
    p_{t,m}(v) = \text{softmax}(\mathbf{h}_{t,m}^\top \mathbf{E}[v]),
\end{equation*}
where $\mathbf{E}[v]$ is the embedding of item $v$. The global reasoning regularization loss is formulated as:
\begin{equation}\label{eq:kl_regularization}
    \mathcal{L}_{\text{KL}} = \frac{1}{TM(TM-1)}\sum_{(t,m)}\sum_{(t',m') \neq (t,m)}\text{KL}(p_{t,m} \| p_{t',m'}),
\end{equation}
where the summation iterates over all pairs of distinct reasoning states across all $T \times M$ positions. This constraint encourages: \textbf{(i)} within each stream, different reasoning steps progressively explore distinct reasoning directions to capture high-order feature interactions; \textbf{(ii)} across different streams, parallel reasoning paths diverge to discover diverse sequence patterns. By penalizing similarity among all reasoning states globally, this mechanism mitigates both intra-stream and inter-stream homogeneity simultaneously.

\subsubsection{\textbf{Reasoning Contrastive Learning}}

To enhance the model's reasoning robustness for better capturing user preferences from sparse interactions, we design a \emph{Reasoning Contrastive Learning (RCL)} objective. Specifically, we introduce dropout strategy along two orthogonal dimensions, representation and interaction, to construct diverse contrastive views.

During the forward pass of the Transformer, we simultaneously apply random dropout to both the hidden representations (with dropout rate $p_{\text{rep}}$) and the attention score matrices (with dropout rate $p_{\text{attn}}$). By performing two independent forward passes with different dropout realizations, we obtain two augmented versions of reasoning outputs: $\{\mathbf{h}_{1,1}^{(1)}, \ldots, \mathbf{h}_{T,M}^{(1)}\}$ and $\{\mathbf{h}_{1,1}^{(2)}, \ldots, \mathbf{h}_{T,M}^{(2)}\}$, where the superscripts denote different dropout masks.

To enforce consistency between the two augmented views, we adopt an in-batch contrastive learning objective~\cite{rusak2024infonce,lin2022improving,hou2023learning}. Specifically, for stream $m$ at reasoning step $t$, we treat the corresponding reasoning states from the two dropout views of the same user as positive pairs, while treating states from other users in the same batch as negative samples. Given a batch of $B$ users, the contrastive loss for stream $m$ at step $t$ is defined as:
\begin{equation*}
\begin{aligned}
    \mathcal{L}_{\text{RCL}}^{t,m} = -\frac{1}{B}\sum_{i=1}^{B}\bigg(&\log\frac{\exp(s_{ii}^{(1,2)}/\tau)}{\sum_{j=1}^{B}\exp(s_{ij}^{(1,2)}/\tau)} 
    + \log\frac{\exp(s_{ii}^{(2,1)}/\tau)}{\sum_{j=1}^{B}\exp(s_{ij}^{(2,1)}/\tau)}\bigg),
\end{aligned}
\end{equation*}
where $s_{ij}^{(1,2)} = \frac{\mathbf{h}_{t,m,i}^{(1)\top}\mathbf{h}_{t,m,j}^{(2)}}{\|\mathbf{h}_{t,m,i}^{(1)}\|\|\mathbf{h}_{t,m,j}^{(2)}\|}$ denotes the cosine similarity between different views (similarly for $s_{ij}^{(2,1)}$), and $\tau$ is the temperature hyperparameter. The overall reasoning contrastive loss aggregates over all reasoning steps and streams:
\begin{equation}\label{eq:contrastive_loss}
    \mathcal{L}_{\text{RCL}} = \frac{1}{TM}\sum_{t=1}^{T}\sum_{m=1}^{M}\mathcal{L}_{\text{RCL}}^{t,m}.
\end{equation}
This objective encourages the model to learn task-relevant representations that remain invariant across different augmented views, thereby enhancing generalization and robustness capabilities.

\subsubsection{\textbf{Mixture-of-Reasoning-Streams Aggregation}}
\label{sec:aggregation}

To synthesize outputs from multiple reasoning streams, naive uniform averaging risks contaminating superior reasoning outputs with inferior ones, leading to suboptimal final representations. To address this, we propose a \emph{Mixture-of-Reasoning-Streams} (MoRS) aggregation method that adaptively weights different streams.

Inspired by gated mechanisms~\cite{qiu2025gated,ma2018modeling,xue2020not}, we introduce a lightweight gating network to compute stream-specific importance weights. Specifically, we feed the encoded representation $\mathbf{h}_0$ (extracted from $\mathbf{H}_{\text{enc}}$) into a gating network:
\begin{equation*}
    \mathbf{g} = \text{softmax}(\mathbf{W}_g\mathbf{h}_0 + \mathbf{b}_g),
\end{equation*}
where $\mathbf{W}_g \in \mathbb{R}^{M \times d}$ and $\mathbf{b}_g \in \mathbb{R}^M$ are learnable parameters, and $\mathbf{g} = [g_1, \ldots, g_M]^\top \in \mathbb{R}^M$ represents the normalized gating weights for $M$ streams. The final output is then computed as:
\begin{equation}\label{eq:slow-thinking-output}
    \mathbf{z}_{\text{rea}} = \sum_{m=1}^{M}g_m\mathbf{z}_m,
\end{equation}
where $\mathbf{z}_m$ is the pooled representation from stream $m$ (see~Eq.\eqref{eq:stream_output}).

\textbf{Dual-Process Inference.} Motivated by the dual-process theory in human cognitive science~\cite{sloman1996empirical,kahneman2011thinking}, which posits that human cognition involves both fast intuitive thinking (System 1) and slow deliberative reasoning (System 2), we combine the encoder output (fast-thinking) with the reasoning output (slow-thinking) during inference. Specifically, the final sequence representation is:
\begin{equation}\label{eq:final_representation}
    \mathbf{z}_{\text{final}} = \mathbf{h}_0 + \mathbf{z}_{\text{rea}},
\end{equation}
where $\mathbf{h}_0$ represents the fast-thinking component directly from the encoder, and $\mathbf{z}_{\text{rea}}$ encapsulates the refined insights from deliberate multi-stream reasoning. This design leverages both rapid pattern recognition and deep logical inference for next-item prediction.

\subsection{Training Objective}\label{sec:training-objective}

The overall training objective combines the \textit{Next-Item Prediction (NIP)} loss (Eq.~\eqref{eq:next-item-prediction}) with the aforementioned global regularization term (Eq.~\eqref{eq:kl_regularization}) and contrastive learning objective (Eq.~\eqref{eq:contrastive_loss}):
\begin{equation}
    \mathcal{L} = \mathcal{L}_{\text{NIP}} + \mathcal{L}_{\text{RCL}} + \lambda\mathcal{L}_{\text{KL}},
\end{equation}
where $\lambda$ is a hyperparameter balancing the regularization strength. Notably, we adopt different strategies for training and inference to fully leverage the dual-process reasoning framework:

\twemoji{fire} \textbf{Learning with Reasoning Output.} During training, we exclusively use the slow-thinking output $\mathbf{z}_{\text{rea}}$ (Eq.~\eqref{eq:slow-thinking-output}) for the next-item prediction task. This design forces the model to develop genuine reasoning capabilities rather than taking shortcuts by relying solely on the fast-thinking encoder output $\mathbf{h}_0$. By preventing direct access to the encoder representation during training, we ensure that the model learns to perform deliberate multi-step reasoning.

\twemoji{snowflake} \textbf{Inference with Dual-Process Integration.} During inference, we combine both the fast-thinking and slow-thinking outputs using $\mathbf{z}_{\text{final}}$ (Eq.~\eqref{eq:final_representation}) to rank candidate items. This dual-process integration leverages the complementary strengths of both reasoning modes: the fast-thinking component $\mathbf{h}_0$ provides rapid pattern recognition based on direct encoding, while the slow-thinking component $\mathbf{z}_{\text{rea}}$ contributes refined insights from deliberate multi-stream reasoning. The synergy between these two modes yields more robust and accurate recommendations.

\section{Theoretical Analysis}

In this section, we provide theoretical foundations to explain why parallel multi-stream reasoning outperforms depth-only baselines. We analyze three perspectives: \textbf{(1)} ensemble theory establishes why diverse reasoning streams reduce prediction error (\S\ref{subsec:ensemble}); \textbf{(2)} dynamical systems analysis reveals the fundamental tension between depth-focused refinement and width-focused diversity (\S\ref{subsec:dynamics}); \textbf{(3)} gating theory explains how adaptive aggregation leverages stream specialization (\S\ref{subsec:gating}). Together, these results provide actionable insights for designing effective parallel reasoning architectures. Complete theoretical proofs are placed in Appendix~\ref{app:proofs}.

\subsection{Why Diversity Reduces Error}
\label{subsec:ensemble}

We first establish the theoretical basis for combining multiple reasoning streams. Let $\hat{p}_m(v|\mathcal{S}_u)$ denote the predicted distribution from stream $m$, and $\bar{p}(v|\mathcal{S}_u) = \frac{1}{M}\sum_{m=1}^M \hat{p}_m(v|\mathcal{S}_u)$ the uniformly weighted ensemble prediction.

\begin{theorem}[Ensemble Error Decomposition]
\label{thm:ensemble}
Define the ensemble loss $\mathcal{L}_{\text{ens}} = \mathbb{E}_{(\mathcal{S}_u,v)\sim\mathcal{D}}[-\log \bar{p}(v|\mathcal{S}_u)]$ and average individual loss $\bar{\mathcal{L}}_{\text{ind}} = \frac{1}{M}\sum_{m=1}^M \mathbb{E}[-\log \hat{p}_m(v|\mathcal{S}_u)]$. Then:
\begin{equation*}
\mathcal{L}_{\text{ens}} \leq \bar{\mathcal{L}}_{\text{ind}} - \mathbb{E}_{\mathcal{S}_u\sim\mathcal{D}}[\mathcal{I}(\mathcal{S}_u)],
\end{equation*}
where the \emph{specialization benefit} $\mathcal{I}(\mathcal{S}_u) \geq 0$ quantifies the gain from diversity, with equality if and only if all $\hat{p}_m$ are identical.
\end{theorem}

The proof (Appendix~\ref{app:proof-ensemble}) leverages Jensen's inequality for the concave logarithm function. Intuitively, when streams produce diverse predictions, the ensemble averages out individual errors, yielding lower loss than the average individual loss.

To connect diversity in representation space to diversity in prediction space, we establish the following result:

\begin{proposition}[Diversity-Specialization Connection]
\label{prop:diversity-spec}
Define representational diversity as $D(\mathcal{S}_u) = \frac{1}{M(M-1)}\sum_{m\neq m'}\|\mathbf{z}_m - \mathbf{z}_{m'}\|^2$. Under linear scoring with bounded item embeddings, the specialization benefit satisfies:
\begin{equation*}
\mathcal{I}(\mathcal{S}_u) \geq c \cdot D(\mathcal{S}_u),
\end{equation*}
where $c > 0$ depends on embedding geometry.
\end{proposition}

This establishes that representational diversity (which our trigger tokens and regularization directly control) translates to prediction diversity, which in turn reduces ensemble error.

\subsection{The Refinement-Diversity Trade-off}
\label{subsec:dynamics}

While Theorem~\ref{thm:ensemble} shows the value of diversity, iterative reasoning introduces a fundamental tension: refinement quality improves with more steps, but diversity may decay. We formalize this trade-off using contraction mapping theory.

\begin{assumption}[Lipschitz Continuity]
\label{assump:lipschitz}
The reasoning function $f$ is $L$-Lipschitz continuous:
\begin{equation*}
\|f(\mathbf{h}, \mathbf{r}; \theta) - f(\mathbf{h}', \mathbf{r}; \theta)\|_2 \leq L\|\mathbf{h} - \mathbf{h}'\|_2.
\end{equation*}
Layer normalization in Transformers naturally constrains output norms and induces contraction behavior. For shallow architectures (e.g., 2-layer Transformers in our implementation), the Lipschitz constant $L$ is typically bounded by a small constant.
\end{assumption}

\begin{theorem}[Diversity Decay Under Iteration]
\label{thm:diversity-decay}
Under Assumption~\ref{assump:lipschitz}, let $D^{(t)} = \frac{1}{M(M-1)}\sum_{m\neq m'}\|\mathbf{h}_{t,m} - \mathbf{h}_{t,m'}\|^2$ denote diversity at reasoning step $t$. Then:
\begin{equation*}
D^{(T)} = L^{2T} D^{(0)} + o(L^{2T}).
\end{equation*}
When $L$ is bounded by a constant smaller than 1, diversity decays exponentially: $D^{(T)} = \exp(-2\gamma T)D^{(0)}$ where $\gamma := -\log L > 0$.
\end{theorem}

The proof (Appendix~\ref{app:proof-dynamics}) follows from iteratively applying the Lipschitz property. This result reveals why depth-only scaling eventually fails: excessive reasoning steps collapse all streams toward a common fixed point, eliminating the diversity benefit.

\begin{corollary}[Refinement-Diversity Trade-off]
\label{cor:tradeoff}
Combining Theorems~\ref{thm:ensemble} and~\ref{thm:diversity-decay}, the ensemble loss at step $T$ satisfies:
\begin{equation*}
\mathcal{L}_{\text{ens}}^{(T)} \leq \bar{\mathcal{L}}_{\text{ind}}^{(T)} - c \cdot e^{-2\gamma T}D^{(0)},
\end{equation*}
where $\bar{\mathcal{L}}_{\text{ind}}^{(T)}$ decreases with $T$ (quality improvement), but the specialization benefit decays exponentially (diversity loss).
\end{corollary}

This formalization explains the empirical over-thinking phenomenon: initially, increasing $T$ reduces $\bar{\mathcal{L}}_{\text{ind}}^{(T)}$ faster than diversity decays; eventually, diminishing returns from refinement are outweighed by diversity loss.

\subsection{Gating Benefits}
\label{subsec:gating}

The preceding analysis assumed uniform weighting. We now characterize the benefit of learned gating and its effect on generalization.

\begin{theorem}[Gating Benefit via Mutual Information]
\label{thm:gating}
Let $\tilde{p}(v|\mathcal{S}_u) = \sum_{m=1}^M w_m(\mathcal{S}_u)\hat{p}_m(v|\mathcal{S}_u)$ denote the gated ensemble. Then:
\begin{equation*}
\mathcal{L}_{\text{gated}} \leq \mathcal{L}_{\text{uniform}} - \mathbb{E}_{\mathcal{S}_u}[I(Z; V|\mathcal{S}_u)],
\end{equation*}
where $Z \sim \mathbf{w}(\mathcal{S}_u)$ is stream selection, $V \sim p^*(v|\mathcal{S}_u)$ is the target, and $I(Z; V|\mathcal{S}_u)$ is conditional mutual information.
\end{theorem}

This shows that gating provides benefit when streams specialize: if different streams excel on different sequence perspectives, the mutual information $I(Z; V|\mathcal{S}_u) > 0$, and adaptive weighting outperforms uniform averaging.




\section{Related Work}

\subsection{Sequential Recommendation}

Sequential recommendation, which aims to predict users' next interactions based on their chronologically ordered behavioral sequences, represents a mainstream paradigm in recommender systems. Early works explored various neural architectures for sequential modeling, including recurrent neural networks~\cite{donkers2017sequential,cui2018mv}, convolutional networks~\cite{tang2018personalized,yan2019cosrec}, and Transformer-based models~\cite{kang2018self,sun2019bert4rec}. SASRec~\cite{kang2018self} pioneered the application of self-attention mechanisms to capture sequential dependencies, while BERT4Rec~\cite{sun2019bert4rec} extended this with bidirectional modeling through masked item prediction. However, these ID-based methods heavily rely on high-quality interaction data and struggle with sparsity and cold-start scenarios.

To address these limitations, recent efforts have incorporated multimodal side information to enrich item representations beyond simple IDs~\cite{hou2022towards,yuan2023go,wei2023multi}. UniSRec~\cite{hou2022towards} learns universal item representations through textual descriptions, enabling cross-domain transfer via unified text-based modeling. MoRec~\cite{yuan2023go} comprehensively investigates the performance gap between multimodal item representations and traditional ID-based sequences, demonstrating the potential of multimodal fusion. Despite these advances, conventional sequential methods still operate under a shallow \emph{input-output} paradigm with fixed computational budgets, limiting their capacity for complex reasoning over user preferences.

\subsection{Reasoning-Enhanced Recommendation}

The remarkable success of large language models (LLMs)~\cite{zhao2023survey,minaee2024large,naveed2025comprehensive}, particularly through Chain-of-Thought (CoT) prompting~\cite{wei2022chain}, has transformed the landscape of AI by shifting from shallow pattern matching to deep reasoning-driven generation. This breakthrough, which extends computation during test time rather than relying solely on encoding capacity, has inspired a new era of reasoning-enhanced recommendation research.

\textbf{Explicit Reasoning Approaches.} The first line of work leverages LLMs' logical capabilities to generate explicit reasoning chains for recommendation tasks. RecGPT-series~\cite{yi2025recgpt-v1,yi2025recgpt-v2} and OneRec-series~\cite{liu2025onerec,zhou2025openonerec} employ LLMs to verbalize insights about user interests, thereby improving prediction accuracy through interpretable reasoning paths. DeepRec~\cite{zheng2025deeprec} introduces multi-turn interactions for iterative refinement of recommendations, while $\text{R}^2$ec~\cite{you2025r} unifies reasoning and recommendation within an autoregressive LLM framework. However, explicit reasoning faces two critical challenges: \textbf{(i)} the absence of well-defined, verifiable reasoning chains in recommendation contexts (unlike mathematical or coding tasks with objective ground truth), and \textbf{(ii)} prohibitive inference latency from verbose token generation, rendering these methods impractical for real-time industrial deployment.

\textbf{Latent Reasoning Approaches.} To overcome these limitations, an emerging paradigm adopts data-driven latent reasoning in continuous representation space, eliminating the need for explicit CoT annotations while maintaining low latency. ReaRec~\cite{tang2025think} pioneers this direction by introducing multi-step autoregressive reasoning with ensemble (ERL) and progressive (PRL) learning strategies. LARES~\cite{liu2025lares} proposes architectural decoupling between encoding (pre-blocks) and reasoning (core-blocks) modules, employing self-supervised pre-training and reinforcement learning for enhanced reasoning quality. OnePiece~\cite{dai2025onepiece} upgrades to block-wise latent reasoning and introduces multi-task progressive supervision to enhance reasoning quality. LatentR$^3$~\cite{zhang2025reinforced} extends this paradigm to LLM backbones with RL-based optimization. Despite their promise, existing latent reasoning methods exclusively focus on \emph{depth-level} computational scaling through sequential reasoning steps, exhibiting diminishing returns and over-thinking issues. In contrast, our work pioneers \emph{width-level} parallel reasoning, exploring multiple diverse reasoning trajectories simultaneously to unlock further performance gains while maintaining real-time efficiency.

\section{Experiments}

In this section, we conduct extensive experiments and analyses to demonstrate the superiority of our proposed PLR framework. We aim to answer the following research questions: \textbf{(RQ1)} How does PLR perform compared to state-of-the-art baselines? \textbf{(RQ2)} What is the contribution of each component in PLR? \textbf{(RQ3)} How do key hyperparameters affect PLR's performance? \textbf{(RQ4)} What insights can we gain from in-depth analysis of PLR's reasoning behavior?

\subsection{Experimental Setup}

\subsubsection{\textbf{Datasets}}

We conduct experiments on three domains from the Amazon Review 2023 dataset~\cite{hou2024bridging}: \textbf{CDs \& Vinyl}, \textbf{Movies \& TV}, and \textbf{Video \& Games}. Following common practice in recommender systems~\cite{tang2025think,zheng2025deeprec}, we treat ratings greater than 3 as positive user interactions. We chronologically split each user's interaction sequence based on the official timestamps, which better reflects real-world industrial scenarios and facilitates fair performance comparison across baselines. For the CDs \& Vinyl and Video \& Games datasets, we filter out users with fewer than 10 interactions. Detailed dataset statistics are presented in Table~\ref{tab:dataset_statistics}.

\noindent
\textbf{Feature Construction:}
For \textit{ID-based} sequential models (\textit{e.g.}, SASRec, BERT4Rec), we construct item representations using the multi-level category hierarchy, store ID, and item ID provided by the dataset. For \textit{text-based} models (\textit{e.g.}, UniSRec), we concatenate the item title, hierarchical categories, and store name to synthesize textual descriptions. These text sequences are then encoded using the BGE embedding model~\cite{chen2024bge} to obtain dense sentence embeddings.

\subsubsection{\textbf{Evaluation Metrics}}

\begin{table}[t]
\centering
\caption{The dataset statistics.}
\vspace{-10pt}
\label{tab:dataset_statistics}
\begin{tabular}{lcccc}
\toprule
\textbf{Dataset} & \textbf{\#User} & \textbf{\#Item} & \textbf{\#Inter.} & \textbf{Sparsity} \\
\midrule
CDs \& Vinyl & 35,238 & 87,969 & 943,399 & 99.97\% \\
Movies \& TV & 51,566 & 101,114 & 1,314,578 & 99.97\% \\
Video \& Games & 90,678 & 22,933 & 728,661 & 99.96\% \\
\bottomrule
\end{tabular}
\vspace{-20pt}
\end{table}

\begin{table*}[t]
\centering
\caption{Performance comparison of different methods on three datasets. The best results are highlighted in \textbf{bold}, and the second-best results are \underline{underlined}. ``Improv.'' denotes the relative improvement of PLR over the best baseline. * indicates statistical significance with $p<0.05$ using paired t-test.}
\label{tab:overall_performance}
\resizebox{\textwidth}{!}{
\begin{tabular}{l|cccc|cccc|cccc}
\toprule
\multirow{2}{*}{\textbf{Method}} & \multicolumn{4}{c|}{\textbf{CDs \& Vinyl}} & \multicolumn{4}{c|}{\textbf{Movies \& TV}} & \multicolumn{4}{c}{\textbf{Video \& Games}} \\
\cmidrule(lr){2-5} \cmidrule(lr){6-9} \cmidrule(lr){10-13}
& R@10 & R@20 & N@10 & N@20 & R@10 & R@20 & N@10 & N@20 & R@10 & R@20 & N@10 & N@20 \\
\midrule
\multicolumn{13}{c}{\cellcolor{cyan!12}\textbf{SASRec}} \\
\cmidrule(lr){1-13}
Base & 0.0544 & 0.0762 & 0.0250 & 0.0305 & 0.0502 & 0.0664 & 0.0230 & 0.0270 & 0.0624 & 0.0937 & 0.0280 & 0.0359 \\
ReaRec-ERL & \underline{0.0572} & \underline{0.0779} & 0.0269 & 0.0321 & 0.0502 & \underline{0.0680} & 0.0237 & \underline{0.0282} & \underline{0.0660} & 0.0953 & \underline{0.0315} & 0.0388 \\
ReaRec-PRL & 0.0553 & 0.0748 & \textbf{0.0273} & \underline{0.0323} & \underline{0.0504} & 0.0650 & \underline{0.0244} & 0.0280 & 0.0630 & \underline{0.1003} & 0.0311 & \underline{0.0405} \\
LARES & 0.0555 & 0.0757 & 0.0260 & 0.0310 & \underline{0.0504} & 0.0645 & 0.0232 & 0.0267 & 0.0648 & 0.0968 & 0.0314 & 0.0395 \\
\rowcolor{gray!20} \textbf{PLR (Ours)} & \textbf{0.0604} & \textbf{0.0873} & \underline{0.0272} & \textbf{0.0339} & \textbf{0.0532} & \textbf{0.0701} & \textbf{0.0250} & \textbf{0.0293} & \textbf{0.0676} & \textbf{0.1033} & \textbf{0.0326} & \textbf{0.0416} \\
\rowcolor{gray!20} Improv. & \imps{5.59} & \imps{12.07} & \impn{0.37} & \imps{4.95} & \imps{5.56} & \imps{3.09} & \imps{2.46} & \imps{3.90} & \imps{2.42} & \imps{2.99} & \imps{3.49} & \imps{2.72} \\
\midrule
\multicolumn{13}{c}{\cellcolor{yellow!12}\textbf{BERT4Rec}} \\
\cmidrule(lr){1-13}
Base & 0.0555 & 0.0745 & 0.0260 & 0.0308 & 0.0444 & 0.0592 & 0.0215 & 0.0253 & 0.0622 & 0.0912 & 0.0302 & 0.0376 \\
ReaRec-ERL & \underline{0.0589} & \underline{0.0793} & \underline{0.0285} & \underline{0.0337} & \underline{0.0518} & \underline{0.0664} & \underline{0.0244} & \underline{0.0281} & \textbf{0.0680} & \underline{0.1017} & \underline{0.0334} & \underline{0.0418} \\
ReaRec-PRL & 0.0541 & 0.0782 & 0.0272 & 0.0333 & 0.0502 & 0.0648 & 0.0235 & 0.0271 & 0.0629 & 0.0989 & 0.0301 & 0.0392 \\
LARES & 0.0513 & 0.0759 & 0.0256 & 0.0318 & 0.0502 & 0.0652 & 0.0232 & 0.0270 & 0.0595 & 0.0960 & 0.0319 & 0.0411 \\
\rowcolor{gray!20} \textbf{PLR (Ours)} & \textbf{0.0626} & \textbf{0.0827} & \textbf{0.0293} & \textbf{0.0344} & \textbf{0.0523} & \textbf{0.0685} & \textbf{0.0247} & \textbf{0.0288} & \underline{0.0656} & \textbf{0.1018} & \textbf{0.0343} & \textbf{0.0435} \\
\rowcolor{gray!20} Improv. & \imps{6.28} & \imps{4.29} & \imps{2.81} & \imps{2.08} & \imp{0.97} & \imps{3.16} & \imp{1.23} & \imps{2.49} & \impn{3.53} & \imp{0.10} & \imps{2.69} & \imps{4.07} \\
\midrule
\multicolumn{13}{c}{\cellcolor{violet!12}\textbf{UniSRec}} \\
\cmidrule(lr){1-13}
Base & 0.0419 & 0.0638 & 0.0195 & 0.0250 & 0.0557 & 0.0724 & 0.0249 & 0.0291 & 0.0656 & 0.1024 & 0.0339 & 0.0432 \\
ReaRec-ERL & \underline{0.0456} & \underline{0.0689} & \underline{0.0208} & \underline{0.0266} & \underline{0.0567} & \underline{0.0782} & 0.0252 & 0.0307 & 0.0673 & 0.1006 & \underline{0.0351} & \underline{0.0435} \\
ReaRec-PRL & 0.0422 & 0.0680 & 0.0195 & 0.0260 & 0.0557 & 0.0754 & 0.0252 & 0.0303 & \underline{0.0688} & \underline{0.1032} & 0.0340 & 0.0427 \\
LARES & 0.0434 & 0.0680 & 0.0194 & 0.0255 & 0.0548 & 0.0765 & \underline{0.0254} & \underline{0.0309} & 0.0687 & 0.1029 & \textbf{0.0353} & \textbf{0.0439} \\
\rowcolor{gray!20} \textbf{PLR (Ours)} & \textbf{0.0524} & \textbf{0.0748} & \textbf{0.0223} & \textbf{0.0279} & \textbf{0.0620} & \textbf{0.0816} & \textbf{0.0271} & \textbf{0.0321} & \textbf{0.0695} & \textbf{0.1054} & 0.0341 & 0.0431 \\
\rowcolor{gray!20} Improv. & \imps{14.91} & \imps{8.56} & \imps{7.21} & \imps{4.89} & \imps{9.35} & \imps{4.35} & \imps{6.69} & \imps{3.88} & \imp{1.02} & \imps{2.13} & \impn{3.40} & \impn{1.82} \\
\bottomrule
\end{tabular}
}
\end{table*}

We adopt two widely-used ranking metrics in the field of recommender systems to evaluate models' personalized modeling capability: \textbf{Normalized Discounted Cumulative Gain at K} (NDCG@K, abbreviated as N@K) and \textbf{Recall at K} (R@K). We report results for $K \in \{10, 20\}$. 
NDCG@K accounts for the position of relevant items in the ranking list with logarithmic discount, while Recall@K measures the proportion of ground-truth items successfully retrieved within the top-K recommendations.

\subsubsection{\textbf{Baselines}}

To comprehensively evaluate PLR's effectiveness, we select three representative backbone architectures and three state-of-the-art latent reasoning-enhanced methods.

\textbf{Base Models.} We choose three widely-adopted sequential recommendation models as our backbone architectures:
\begin{itemize}[leftmargin=*,noitemsep,topsep=2pt]
    \item \textbf{SASRec}~\cite{kang2018self}: A pioneering Transformer-based sequential recommendation model that employs unidirectional multi-head self-attention to capture sequential dependencies in user behavior.
    \item \textbf{BERT4Rec}~\cite{sun2019bert4rec}: An extension of SASRec that leverages bidirectional self-attention with Cloze task (masked item prediction) for sequence modeling, enabling richer contextual representations.
    \item \textbf{UniSRec}~\cite{hou2022towards}: A text-based sequential recommender that learns universal item representations through textual features and Mixture-of-Experts (MoE) adapters for cross-domain transfer.
\end{itemize}

\textbf{Latent Reasoning-Enhanced Methods.} We focus exclusively on latent reasoning baselines to ensure fair comparison under comparable inference latency and model capacity. Explicit reasoning approaches and LLM-based methods are excluded due to their prohibitive computational overhead for real-time online serving and vastly different parameter scales.
\begin{itemize}[leftmargin=*,noitemsep,topsep=2pt]
    \item \textbf{ReaRec-ERL}~\cite{tang2025think}: The first latent reasoning framework for sequential recommendation, which firstly introduces multi-step autoregressive computation in latent space to explore test-time reasoning. ERL (Ensemble Reasoning Learning) aggregates outputs from all reasoning steps via average pooling.
    \item \textbf{ReaRec-PRL}~\cite{tang2025think}: A variant of ReaRec that employs PRL (Progressive Reasoning Learning), which incrementally refines representations and uses only the final step's output for prediction.
    \item \textbf{LARES}~\cite{liu2025lares}: A two-stage reasoning architecture consisting of pre-blocks for encoding and core-blocks for reasoning. LARES incorporates self-supervised pre-training and reinforcement learning-based post-training to enhance reasoning capabilities.
\end{itemize}

We implement different latent reasoning methods on top of the three backbone architectures for comprehensive evaluation.

\begin{figure*}[t]
  \centering
  \includegraphics[width=\linewidth]{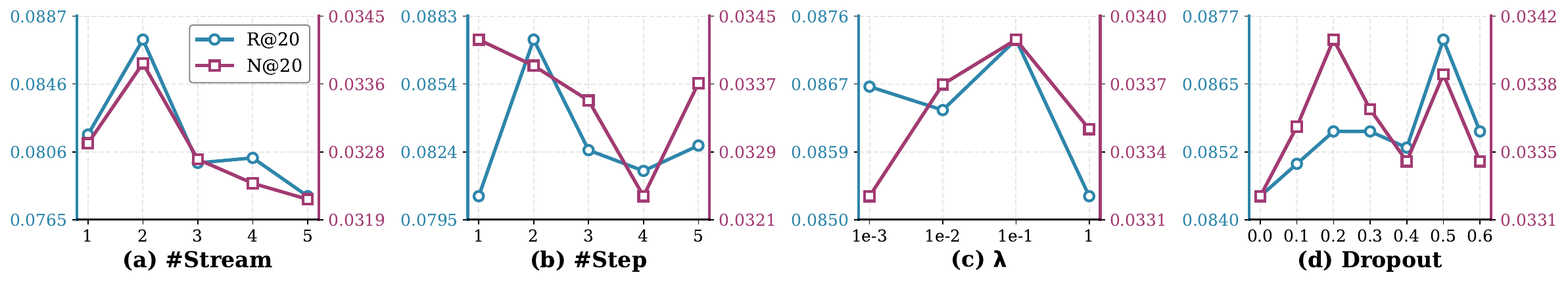}
  \vspace{-20pt}
  \caption{Parameter sensitivity analysis on the CDs \& Vinyl dataset with SASRec backbone.}
  \label{fig:param}
\end{figure*}

\subsubsection{\textbf{Implementation Details}}

All experiments are conducted using the PyTorch framework on NVIDIA A100 GPUs. To ensure fair comparison across all models, we standardize the embedding dimension to $d=256$ and batch size to $B=2048$. We employ the Adam optimizer~\cite{kingma2014adam} with a learning rate of $1e^{-3}$ and apply the GeLU activation function. User interaction sequences are truncated to a maximum length of 50 items. For text-based models, we apply Principal Component Analysis (PCA) to reduce the BGE-encoded representations from their original dimensionality to 768 dimensions, preserving the most informative features.

For PLR, we perform grid search to determine optimal hyperparameters: the number of reasoning streams $M$ and reasoning steps $T$ are searched in $\{1, 2, 3, 4, 5\}$; the KL regularization weight $\lambda$ is tuned in $\{1e^{-3}, 1e^{-2}, 1e^{-1}, 1\}$; dropout rates $p_{\text{rep}}$ and $p_{\text{attn}}$ are set equal and searched in $\{0.1, 0.2, 0.3, 0.4, 0.5, 0.6\}$; and the temperature $\tau$ is fixed at 1. We use early stopping based on validation NDCG@10 with a patience of 10 epochs. To optimize inference efficiency, we employ \textbf{\textit{KV Caching}} technique that stores and reuses the computed key-value pairs from the input sequence encoding across all reasoning streams and steps, significantly reducing redundant computation overhead during multi-stream parallel reasoning.

\subsection{Overall Performance (RQ1)}

Table~\ref{tab:overall_performance} presents the overall performance comparison across three datasets. We highlight several key observations as follows:

\textbf{(1)} Reasoning-enhanced methods outperform base models across most settings, validating the effectiveness of test-time computational scaling. On CDs \& Vinyl with SASRec, ReaRec-ERL improves Recall@10 by 5.15\% and Recall@20 by 2.23\%. On Movies \& TV with BERT4Rec, ReaRec-ERL demonstrates strong improvements of 16.67\% in Recall@10 and 13.49\% in NDCG@10. These gains validate that allocating additional computation during inference enables deeper reasoning, particularly beneficial for sparse datasets.

\textbf{(2)} Among depth-only reasoning methods, ReaRec-ERL generally demonstrates the strongest performance across most scenarios. For instance, on CDs \& Vinyl with SASRec, ReaRec-ERL achieves Recall@10 of 0.0572 and NDCG@10 of 0.0269, outperforming both ReaRec-PRL and LARES. 
However, the performance differences among these methods remain relatively modest, indicating inherent limitations in depth-only reasoning where simply increasing reasoning depth yields diminishing returns.

\textbf{(3)} PLR achieves substantial improvements over all baselines across the majority of metrics. On CDs \& Vinyl with SASRec, PLR attains 12.07\% improvement in Recall@20 over the best baseline ReaRec-ERL, with even more remarkable gains of 14.91\% in Recall@10 on UniSRec. While improvements vary across different datasets and metrics, PLR demonstrates superior performance in most scenarios, particularly excelling on sparser datasets like CDs \& Vinyl where it achieves over 10\% improvements. These results demonstrate that width-level parallel reasoning effectively captures diverse user interests beyond single-trajectory depth reasoning. Crucially, these gains come with minimal inference overhead (Table~\ref{tab:efficiency}), making PLR practical for real-time systems.

\begin{table}[t]
\centering
\caption{Ablation study. The experiments are conducted based on SASRec on the CDs \& Vinyl and Video \& Games datasets.}
\label{tab:ablation}
\begin{tabular}{lcccccc}
\toprule
\multirow{2}{*}{\textbf{Methods}} & \multicolumn{2}{c}{\textbf{CDs \& Vinyl}} & & \multicolumn{2}{c}{\textbf{Video \& Games}} \\
\cmidrule{2-3} \cmidrule{5-6}
& R@20 & N@20 & & R@20 & N@20 \\
\midrule
w/o MoRS & 0.0785 & 0.0321 & & 0.0997 & 0.0415 \\
w/o RCL & 0.0782 & 0.0322 & & 0.0970 & 0.0387 \\
w/o KL & 0.0853 & 0.0334 & & 0.0979 & 0.0380 \\
\midrule
\textbf{PLR (Full)} & \textbf{0.0873} & \textbf{0.0339} & & \textbf{0.1033} & \textbf{0.0416} \\
\bottomrule
\end{tabular}
\end{table}

\subsection{Ablation Study (RQ2)}

To validate the effectiveness of each component in PLR, we conduct ablation experiments on two representative datasets (CDs \& Vinyl and Video \& Games) with SASRec backbone. Table~\ref{tab:ablation} presents the results of four ablated variants: 
\textbf{(i) w/o MoRS} replaces the adaptive mixture-of-reasoning-streams aggregation with uniform average pooling; \textbf{(ii) w/o RCL} removes the reasoning contrastive learning objective; and \textbf{(iii) w/o KL} eliminates the global regularization.

The results demonstrate that each component contributes meaningfully to PLR's performance. Removing MoRS leads to notable performance degradation (Recall@20 drops from 0.0873 to 0.0785 on CDs \& Vinyl), validating that adaptive gating effectively identifies superior reasoning streams, consistent with Theorem~\ref{thm:gating}. Removing RCL causes decline particularly on Video \& Games (Recall@20 drops from 0.1033 to 0.0970), confirming its role in enhancing robustness. Removing KL also leads to performance drops on both datasets, though the degradation is relatively modest, suggesting that diversity regularization provides consistent but moderate benefits. Overall, all components show positive contributions, with MoRS and RCL demonstrating more substantial effects.

\subsection{Sensitivity Analysis (RQ3)}

We conduct sensitivity analysis on key hyperparameters to understand their impact on PLR's performance. Figure~\ref{fig:param} presents the results on the CDs \& Vinyl dataset with SASRec backbone, measured by NDCG@20 and Recall@20.

\textbf{Number of Reasoning Streams ($M$).} As shown in Figure~\ref{fig:param}(a), performance peaks at $M=2$, achieving optimal NDCG@20 and Recall@20. Increasing $M$ beyond 2 leads to performance degradation, likely because excessive streams introduce redundant reasoning paths that dilute superior streams' contributions, and limited training data becomes insufficient for effectively optimizing numerous specialized streams. We set $M=2$ as the default configuration.

\textbf{Number of Reasoning Steps ($T$).} Figure~\ref{fig:param}(b) shows performance peaks at $T=2$, then declines with deeper reasoning. This aligns with our theoretical analysis (Theorem~\ref{thm:diversity-decay}): while moderate reasoning steps enable sufficient refinement, excessive depth causes diversity decay due to Lipschitz contraction. The optimal $T=2$ balances refinement quality and diversity preservation, confirming the over-thinking phenomenon in depth-only methods.

\textbf{Regularization Weight ($\lambda$).} Figure~\ref{fig:param}(c) shows optimal performance at $\lambda=0.1$. Insufficient regularization ($\lambda<0.1$) fails to enforce diversity, allowing streams to collapse into homogeneous patterns. Overly aggressive regularization ($\lambda>0.1$) negatively impacts learning by pushing different reasoning streams too far apart, hindering convergence toward high-quality solutions.

\textbf{Dropout Rate ($p$).} Figure~\ref{fig:param}(d) indicates that moderate dropout rates $p \in [0.2, 0.5]$ yield the best performance. Low dropout ($p<0.2$) provides insufficient augmentation for contrastive learning, while excessive dropout ($p>0.5$) corrupts reasoning by removing too much information. The optimal range effectively balances augmentation diversity and information preservation.

\begin{figure}[t]
  \centering
  \includegraphics[width=\linewidth]{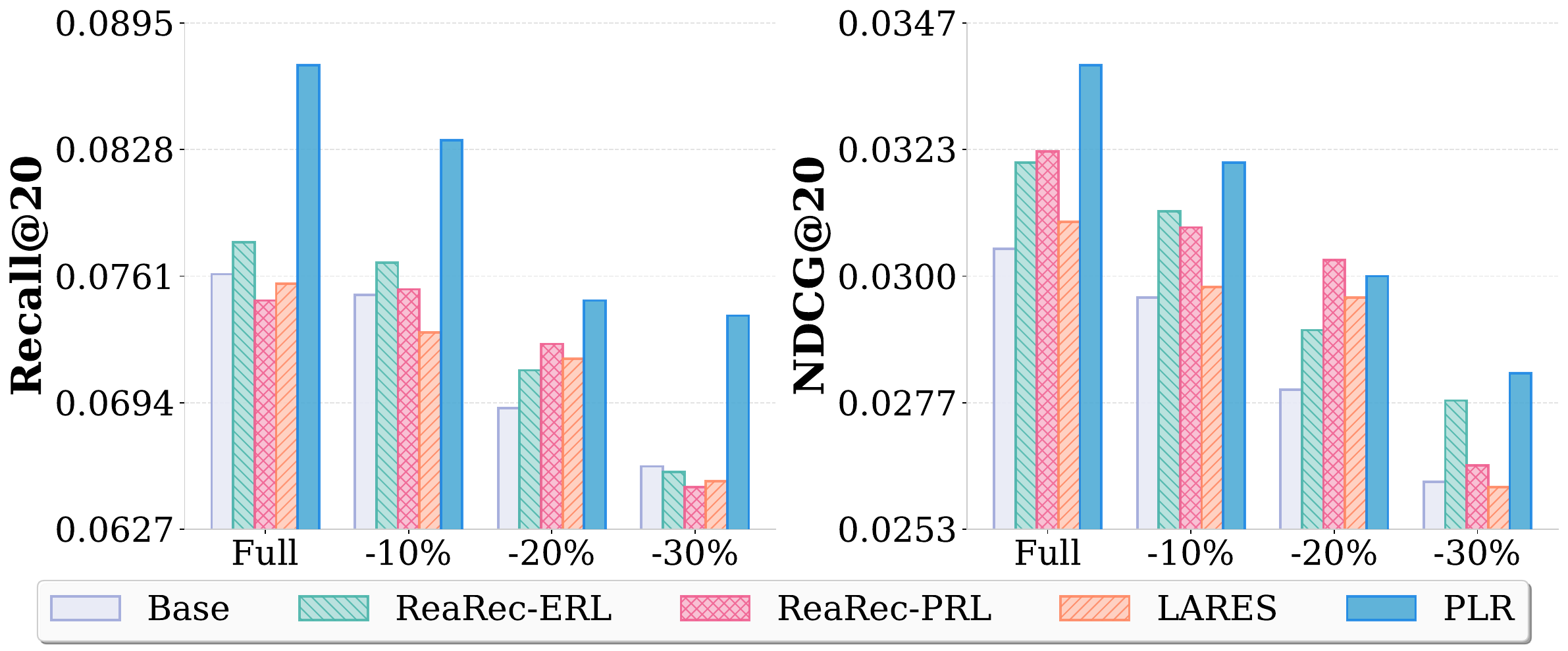}
  \vspace{-20pt}
  \caption{Robustness analysis on the CDs \& Vinyl dataset.}
  \label{fig:robust}
  \vspace{-10pt}
\end{figure}

\subsection{In-Depth Analysis (RQ4)}

\subsubsection{\textbf{Robustness Analysis}}
To evaluate model robustness under extreme data sparsity, we randomly remove 10\%, 20\%, and 30\% of interactions from user sequences in the CDs \& Vinyl test set. Figure~\ref{fig:robust} shows that while all models exhibit performance decline as sparsity increases, reasoning-enhanced methods demonstrate superior robustness compared to base models. At 20\% missing rate, the base SASRec's performance drops significantly, whereas reasoning methods maintain relatively stable performance. Among all approaches, PLR exhibits the strongest robustness: at 30\% missing rate, PLR maintains NDCG@20 around 0.028, substantially outperforming other methods. This superior robustness stems from PLR's parallel reasoning mechanism, where multiple reasoning streams enable the model to explore diverse pathways and compensate for lost behavioral signals when certain interaction patterns become unavailable. These results demonstrate PLR's practical value for real-world recommendation scenarios where user behavioral data is inherently sparse and incomplete.

\begin{figure}[ht]
  \centering
  \includegraphics[width=\linewidth]{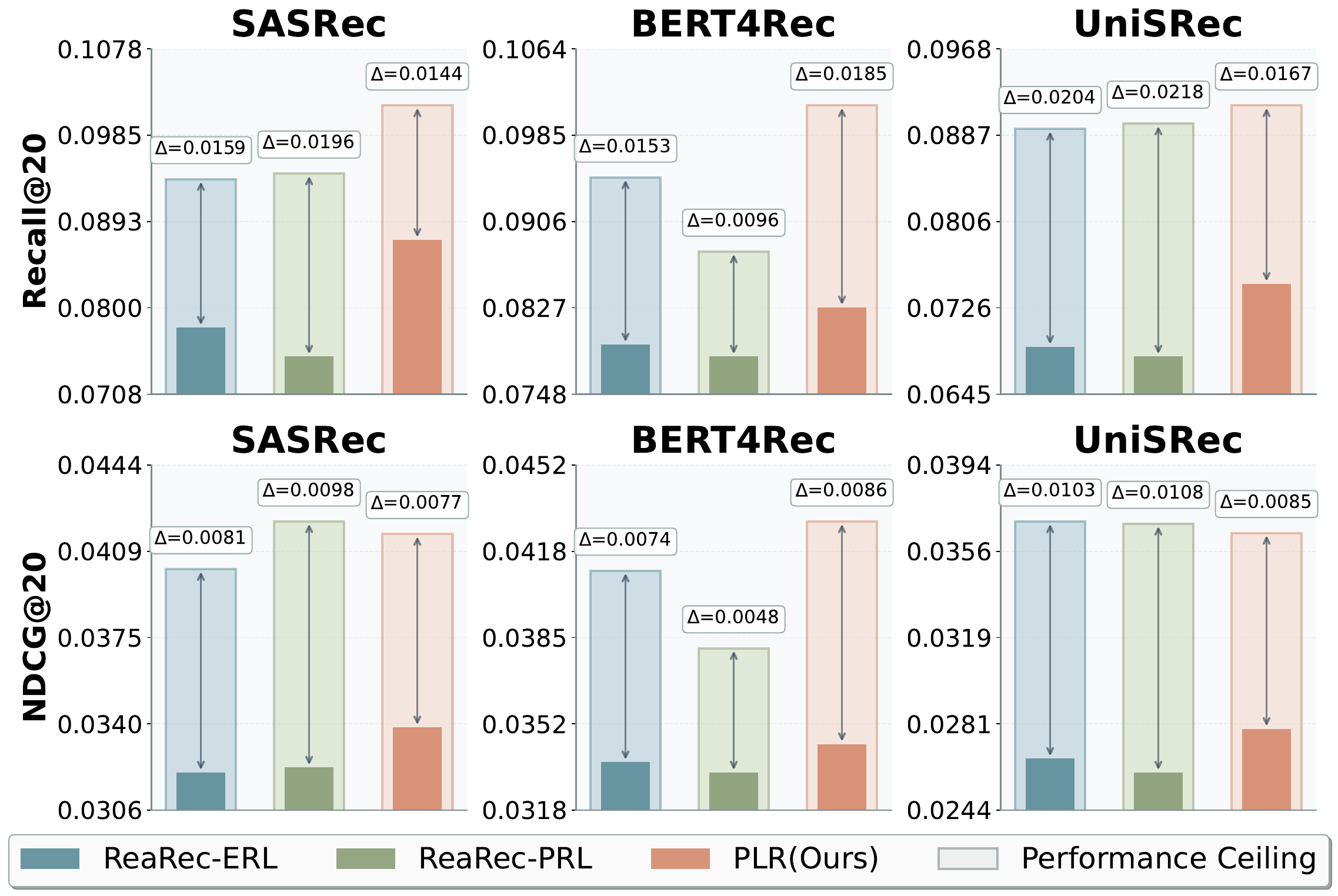}
  \vspace{-20pt}
  \caption{Performance ceiling analysis on the CDs \& Vinyl dataset with different backbones.}
  \label{fig:ceiling}
  \vspace{-10pt}
\end{figure}

\subsubsection{\textbf{Performance Ceiling Analysis}}
To understand the theoretical potential of different reasoning approaches, we analyze performance ceilings by evaluating each reasoning step's output independently and selecting the best-performing step for each user sample (oracle selection). Figure~\ref{fig:ceiling} compares current performance (using fixed aggregation strategies) against oracle ceiling performance across three backbones on the CDs \& Vinyl dataset. The results reveal that PLR and ReaRec-series achieve comparable oracle ceilings (\textit{e.g.}, $\text{NDCG@20}\in[0.0402,0.0416]$), indicating similar theoretical upper bounds. However, PLR exhibits notably smaller gaps between current and oracle performance, suggesting that its parallel multi-stream design enables more effective utilization of reasoning capacity. In contrast, depth-only methods show larger performance gaps, indicating that appropriate step selection is more critical but challenging without width-level exploration. This demonstrates that PLR's co-design of depth and width provides a more accessible path to near-optimal performance, reducing reliance on oracle-level step selection while maintaining competitive ceiling potential.

\begin{table}[t]
\centering
\caption{Efficiency Comparison of Different Methods}
\vspace{-10pt}
\label{tab:efficiency}
\begin{tabular}{l|ccc}
\toprule
\textbf{Metric} & \textbf{SASRec} & \textbf{ReaRec} & \textbf{PLR} \\
\midrule
FLOPs ($\times 10^8$) & 1.3448 & 1.3810 & 1.4150 \\
Improv.  & - & +2.69\% & +5.22\% \\
Latency (s) & 0.7844 & 0.8279 & 0.8299 \\
Cost Inc. & - & +5.55\% & +5.80\% \\
\bottomrule
\end{tabular}
\vspace{-10pt}
\end{table}

\subsubsection{\textbf{Efficiency Analysis}}

To evaluate the computational overhead, we compare the FLOPs and inference latency per sample on the CDs \& Vinyl dataset with SASRec backbone. As shown in Table~\ref{tab:efficiency}, PLR introduces 5.22\% additional FLOPs and 5.80\% latency increase compared to the base SASRec, which represents a remarkably efficient trade-off considering the substantial performance gains. Notably, PLR's computational overhead remains comparable to ReaRec (2.69\% FLOPs and 5.55\% latency increase), despite exploring multiple reasoning streams simultaneously. This efficiency stems from two key design choices: \textbf{(i)} KV Caching mechanism that reuses the shared encoded key-value pairs from the input and prior reasoning states across all streams and steps, avoiding redundant computation; \textbf{(ii)} vectorized parallel multi-stream reasoning, which leverages modern GPU parallelism to process $M$ streams simultaneously without proportional latency scaling. Consequently, PLR achieves width-level computational scaling while maintaining real-time inference feasibility, making it practical for industrial deployment where both accuracy and response time are critical.

\begin{figure}[t]
  \centering
  \includegraphics[width=\linewidth]{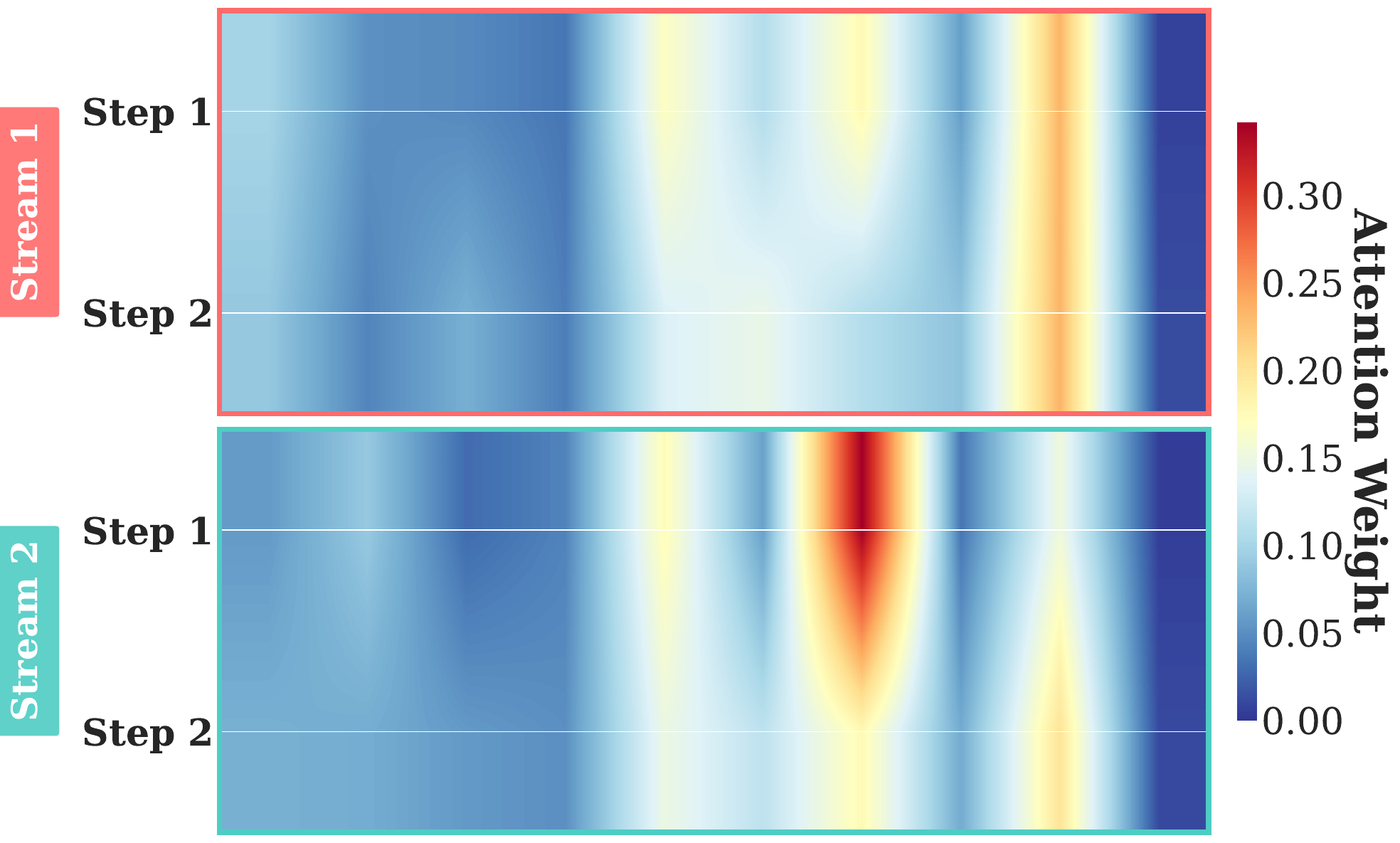}
  \vspace{-20pt}
  \caption{Attention visualization illustration.}
  \label{fig:attention}
  \vspace{-10pt}
\end{figure}

\subsubsection{\textbf{Attention Visualization Analysis}}
To provide deeper insights into the multi-stream reasoning mechanism, we visualize the attention distributions across different streams and reasoning steps in Figure~\ref{fig:attention}. The visualization reveals that different streams exhibit distinct attention patterns, focusing on complementary aspects of the input sequence: while one stream concentrates on recent interactions, another explores long-term dependencies. This diversity in attention allocation substantiates that our multi-stream architecture encourages exploration of heterogeneous reasoning paths, enabling more comprehensive coverage of user interests and contributing to PLR's superior performance.

\section{Conclusion}

In this work, we propose Parallel Latent Reasoning (PLR), a novel framework that pioneers width-level computational scaling for sequential recommendation. Unlike existing depth-only reasoning methods that suffer from diminishing returns, PLR explores multiple diverse reasoning trajectories simultaneously through learnable trigger tokens, global reasoning regularization, and adaptive mixture-of-reasoning-streams aggregation. Extensive experiments on three real-world datasets demonstrate that PLR substantially outperforms state-of-the-art baselines while maintaining real-time inference efficiency. Theoretical analysis further validates the effectiveness of parallel reasoning in reducing prediction error through ensemble diversity and mitigating over-thinking issues. Our work opens new avenues for reasoning-enhanced recommendation by demonstrating that width-level scaling offers a complementary dimension to depth scaling for unlocking models' reasoning potential.

Looking forward, we plan to further explore synergistic integration strategies that jointly optimize both parallel and depth reasoning scaling, potentially enabling more sophisticated multi-dimensional reasoning architectures.



\bibliographystyle{ACM-Reference-Format}
\bibliography{sample-base}

\appendix

\section{Complete Theoretical Proofs}
\label{app:proofs}

\subsection{Proof of Theorem~\ref{thm:ensemble} (Ensemble Error Decomposition)}
\label{app:proof-ensemble}

\begin{proof}
For any sequence $\mathcal{S}_u$, the expected negative log-likelihood under $p^*(v|\mathcal{S}_u)$ is:
\begin{equation*}
\mathbb{E}_{v\sim p^*}[-\log \bar{p}(v|\mathcal{S}_u)] = -\sum_v p^*(v|\mathcal{S}_u) \log\left(\frac{1}{M}\sum_{m=1}^M \hat{p}_m(v|\mathcal{S}_u)\right).
\end{equation*}
By Jensen's inequality (concavity of $\log$):
\begin{equation*}
-\log\left(\frac{1}{M}\sum_{m=1}^M \hat{p}_m(v|\mathcal{S}_u)\right) \leq -\frac{1}{M}\sum_{m=1}^M \log \hat{p}_m(v|\mathcal{S}_u),
\end{equation*}
with strict inequality when predictions differ. Therefore:
\begin{align*}
\mathbb{E}_{v\sim p^*}[-\log \bar{p}(v|\mathcal{S}_u)] &\leq \frac{1}{M}\sum_{m=1}^M \mathbb{E}_{v\sim p^*}[-\log \hat{p}_m(v|\mathcal{S}_u)] \\
&= \frac{1}{M}\sum_{m=1}^M \mathcal{L}_m(\mathcal{S}_u),
\end{align*}
where $\mathcal{L}_m(\mathcal{S}_u)$ is the loss of stream $m$ on sequence $\mathcal{S}_u$. Taking expectation over $\mathcal{S}_u \sim \mathcal{D}$ yields:
\begin{equation*}
\mathcal{L}_{\text{ens}} \leq \bar{\mathcal{L}}_{\text{ind}}.
\end{equation*}
The gap $\mathcal{I}(\mathcal{S}_u) = \bar{\mathcal{L}}_{\text{ind}} - \mathcal{L}_{\text{ens}}(\mathcal{S}_u)$ quantifies the benefit of diverse predictions. By the AM-GM inequality, $\mathcal{I}(\mathcal{S}_u) \geq 0$ with equality if and only if all $\hat{p}_m(v|\mathcal{S}_u)$ are identical.
\end{proof}

\subsection{Proof of Proposition~\ref{prop:diversity-spec} (Diversity-Specialization Connection)}
\label{app:proof-diversity-spec}

\begin{proof}
Under linear scoring, $\hat{p}_m(v|\mathcal{S}_u) = \text{softmax}(\langle \mathbf{z}_m, \mathbf{e}_v\rangle)$. For two distinct streams $m, m'$:
\begin{equation*}
\|\hat{p}_m(\cdot|\mathcal{S}_u) - \hat{p}_{m'}(\cdot|\mathcal{S}_u)\|_{\text{TV}} \geq \frac{\delta_e}{2R}\|\mathbf{z}_m - \mathbf{z}_{m'}\|,
\end{equation*}
where $\|\cdot\|_{\text{TV}}$ is total variation distance, $R$ bounds item embeddings, and $\delta_e$ is minimum embedding separation. By Pinsker's inequality, total variation bounds KL divergence:
\begin{equation*}
\text{KL}(\hat{p}_m \| \hat{p}_{m'}) \geq \frac{1}{2}\|\hat{p}_m - \hat{p}_{m'}\|_{\text{TV}}^2 \geq \frac{\delta_e^2}{8R^2}\|\mathbf{z}_m - \mathbf{z}_{m'}\|^2.
\end{equation*}
The specialization benefit (from Jensen's inequality gap) is lower bounded by the average pairwise KL divergence:
\begin{equation*}
\mathcal{I}(\mathcal{S}_u) \geq \frac{1}{M(M-1)}\sum_{m\neq m'}\text{KL}(\hat{p}_m \| \hat{p}_{m'}) \geq \frac{\delta_e^2}{8R^2} D(\mathcal{S}_u).
\end{equation*}
Setting $c = \delta_e^2/(8R^2) > 0$ completes the proof.
\end{proof}

\subsection{Proof of Theorem~\ref{thm:diversity-decay} (Diversity Decay)}
\label{app:proof-dynamics}

\begin{proof}
By the Lipschitz property (Assumption~\ref{assump:lipschitz}):
\begin{equation*}
\|\mathbf{h}_{t,m} - \mathbf{h}_{t,m'}\| = \|f(\mathbf{h}_{t-1,m}, \mathbf{r}_t) - f(\mathbf{h}_{t-1,m'}, \mathbf{r}_t)\| \leq L\|\mathbf{h}_{t-1,m} - \mathbf{h}_{t-1,m'}\|.
\end{equation*}
Iterating from $t=0$ to $t=T$:
\begin{equation*}
\|\mathbf{h}_{T,m} - \mathbf{h}_{T,m'}\| \leq L^T\|\mathbf{h}_{0,m} - \mathbf{h}_{0,m'}\| = L^T\|\boldsymbol{\tau}_m - \boldsymbol{\tau}_{m'}\|.
\end{equation*}
Squaring both sides:
\begin{equation*}
\|\mathbf{h}_{T,m} - \mathbf{h}_{T,m'}\|^2 \leq L^{2T}\|\boldsymbol{\tau}_m - \boldsymbol{\tau}_{m'}\|^2.
\end{equation*}
Averaging over all pairs $(m, m')$:
\begin{equation*}
D^{(T)} = \frac{1}{M(M-1)}\sum_{m\neq m'}\|\mathbf{h}_{T,m} - \mathbf{h}_{T,m'}\|^2 \leq L^{2T}D^{(0)}.
\end{equation*}
For $L < 1$, writing $L = e^{-\gamma}$ with $\gamma = -\log L > 0$ yields the exponential form:
\begin{equation*}
D^{(T)} \leq e^{-2\gamma T}D^{(0)}.
\end{equation*}
\end{proof}

\subsection{Proof of Theorem~\ref{thm:gating} (Gating Benefit)}
\label{app:proof-gating}

\begin{proof}
For any $\mathcal{S}_u$, by Jensen's inequality:
\begin{align*}
\mathbb{E}_{v\sim p^*}[-\log \tilde{p}(v|\mathcal{S}_u)] &= -\sum_v p^*(v|\mathcal{S}_u) \log\left(\sum_{m=1}^M w_m(\mathcal{S}_u)\hat{p}_m(v|\mathcal{S}_u)\right) \\
&\leq -\sum_v p^*(v|\mathcal{S}_u) \sum_{m=1}^M w_m(\mathcal{S}_u) \log \hat{p}_m(v|\mathcal{S}_u) \\
&= \sum_{m=1}^M w_m(\mathcal{S}_u) \mathbb{E}_{v\sim p^*}[-\log \hat{p}_m(v|\mathcal{S}_u)].
\end{align*}
The gap between gated and uniform ensemble is:
\begin{align*}
\mathcal{L}_{\text{uniform}}(\mathcal{S}_u) - \mathcal{L}_{\text{gated}}(\mathcal{S}_u) &= \mathbb{E}_{v\sim p^*}\left[\log \tilde{p}(v|\mathcal{S}_u) - \log \bar{p}(v|\mathcal{S}_u)\right].
\end{align*}
When streams specialize (per-stream losses vary across $m$ and $\mathcal{S}_u$), optimal weighting assigns higher $w_m(\mathcal{S}_u)$ to streams with lower losses. This gain is quantified by the conditional mutual information:
\begin{equation*}
I(Z; V|\mathcal{S}_u) = \sum_{m=1}^M w_m(\mathcal{S}_u) \mathbb{E}_{v\sim p^*}\left[\log \frac{\hat{p}_m(v|\mathcal{S}_u)}{\tilde{p}(v|\mathcal{S}_u)}\right],
\end{equation*}
which measures how much knowledge of the selected stream $Z$ reveals about the target $V$. Under specialization, $I(Z; V|\mathcal{S}_u) > 0$.
\end{proof}



\end{document}